\documentclass[acmsmall, nonacm]{acmart}

\usepackage{pifont}
\usepackage{tcolorbox}
\usepackage{xcolor}
\usepackage{listings}
\usepackage{enumitem}
\usepackage{algorithm}
\usepackage{graphicx}
\usepackage{natbib}
\usepackage{url}

\begin{document}

\title{Investigating Sharding Advancements, Methodologies, and Adoption Potential in Hedera}

\fancyhead{}
\author{Ziwei Wang}
\affiliation{
  \institution{Wuhan University}
  \country{China}
}
\email{t4stek1ng@whu.edu.cn}

\author{Cong Wu}
\affiliation{
  \institution{Exponential Science}
  \country{United Kingdom}
}
\affiliation{
  \institution{The University of Hong Kong}
  \country{China}
}
\email{cnacwu@hku.edu.cn}

\author{Paolo Tasca}
\affiliation{
  \institution{Exponential Science}
  \country{United Kingdom}
}
\email{p.tasca@exp.science}
\begin{abstract}

Sharding has emerged as a critical solution to address the scalability challenges faced by blockchain networks, enabling them to achieve higher transaction throughput, reduced latency, and optimized resource usage. This paper investigates the advancements, methodologies, and adoption potential of sharding in the context of Hedera, a distributed ledger technology known for its unique Gossip about Gossip protocol and asynchronous Byzantine Fault Tolerance (ABFT). We explore various academic and industrial sharding techniques, emphasizing their benefits and trade-offs. Building on these insights, we propose a hybrid sharding solution for Hedera that partitions the network into local and global committees, facilitating efficient cross-shard transactions and ensuring robust security through dynamic reconfiguration. Our analysis highlights significant reductions in storage and communication overhead, improved scalability, and enhanced fault tolerance, demonstrating the feasibility and advantages of integrating sharding into Hedera's architecture.

\end{abstract}



\maketitle

\section{Introduction}

Blockchain technology, despite its transformative potential in sectors such as finance, healthcare, and supply chain management, is hindered by scalability problems~\cite{varma2019blockchain,attaran2022blockchain,dutta2020blockchain,yang2020review}. These limitations manifest themselves in low transaction throughput and high latency, making blockchains such as Bitcoin and Ethereum under-performing compared to traditional systems~\cite{li2023survey}. Sharding, a concept borrowed from distributed databases, has emerged as a promising solution to address these challenges~\cite{bagui2015database}. By partitioning the blockchain network into smaller, parallelizable units called shards, the technology enables the simultaneous processing of transactions and data storage, significantly enhancing scalability and performance.

The benefits of sharding are multiple. One key point is that it increases transaction throughput by allowing multiple shards to process transactions in parallel, directly reducing network congestion~\cite{dang2019towards}. In addition, it is essential to note that sharding reduces the computational and storage burden on individual nodes, making the network more accessible and decentralized~\cite{wang2019sok}. These improvements collectively address the scalability trilemma, enabling blockchains to achieve security, decentralization, and scalability simultaneously.

\section{Background}

\subsection{Sharding preliminary}

Traditional blockchains process transactions sequentially and each node maintains a complete copy of the blockchain~\cite{gupta2021blockchain}, leading to scalability limitations. The primary goal of sharding is to achieve horizontal scalability, where the network's capacity increases as more nodes are added. By allowing nodes to work collaboratively on smaller portions of the blockchain, sharding helps address the scalability trilemma, which posits that blockchains cannot simultaneously achieve decentralization, scalability, and security~\cite{werth2023review}. Let us observe how the sharding impacts those three properties:

\begin{itemize}

\item \textbf{Decentralization. }
Effective sharding implementations include decentralized methods for cross-shard communication, ensuring that transactions spanning multiple shards are processed without relying on a central authority~\cite{liu2022building}. This maintains the decentralized ethos of blockchain networks. In addition, randomly assigning nodes to different shards prevents any single entity from gaining control over a shard, thus preserving the overall decentralization of the network~\cite{xu2020flexible}.

\item \textbf{Scalability. }
Sharding involves dividing both the workload and the state of the blockchain, enabling multiple shards to process transactions simultaneously~\cite{li2022achieving}. This reduces the computational and storage burden on individual nodes, enhancing overall system efficiency without sacrificing decentralization or security.

\item \textbf{Security. }
To ensure robustness, the sharding systems incorporate various mechanisms.

\begin{itemize}
    \item \textbf{Random Node Assignment: }Randomly assigning nodes prevents adversaries from concentrating malicious nodes within a single shard. The tradoff of this method leads to a high cross-shard ratio and reduced throughput~\cite{tao2024throughput}.

    \item \textbf{Cross-Shard Communication: }In a sharded blockchain, each shard processes its own transactions independently, but some operations require coordination or data sharing between different shards. This is where cross-shard communication comes in. Protocols such as atomic commit and two-phase locking ensure secure and efficient handling of cross-shard transactions.

    \item \textbf{Shard Reconfiguration: } Due to the risk of a corruption attack by an adversary, the nodes in the shards must be updated regularly~\cite{liu2020secure}. During the node selection and reconfiguration process of the committees, a complex secure randomness generation protocol is required.
\end{itemize}
\end{itemize}
\begin{figure*}[ht]
	\centering
	\includegraphics[scale=0.9]{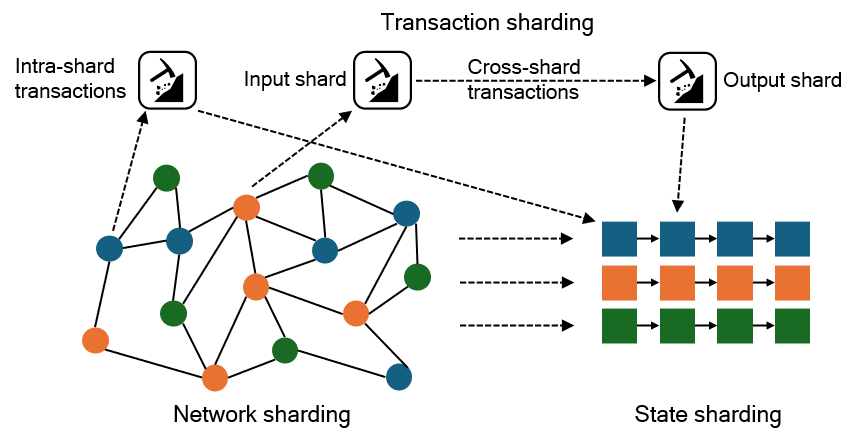}
	\caption{Schema of three types of sharding}
	\label{fig:taxonomy}
\end{figure*}

\subsection{Taxonomy of Sharding Techniques}

Sharding in blockchain can be categorized into three main types based on the components it addresses: network sharding, transaction sharding, and state sharding. These types are not mutually exclusive and can be combined to achieve comprehensive scalability improvements~\cite{liu2022building}.

\textbf{Network Sharding. }
In network sharding, the blockchain's peer-to-peer network is divided into smaller groups or committees, called shards. Each shard handles a portion of the network's communication and verification tasks. This reduces the communication overhead for individual nodes, as they only need to process data relevant to their shard. It is commonly used as the foundational layer for other sharding types, especially in public blockchains like Ethereum.

\textbf{Transaction Sharding. }
This technique partitions transactions into disjoint subsets, with each shard responsible for processing only the transactions assigned to it. Shards execute intra-shard transactions independently while collaborating on cross-shard transactions. This minimizes the computational workload of the nodes by reducing the number of transactions they need to validate. The challenge here is that cross-shard transaction processing can introduce complexity and latency.

\textbf{State Sharding. }
In state sharding, the blockchain's entire state, including account balances and smart contracts, is divided among shards. Each shard stores and processes only its assigned portion of the state. This reduces storage requirements for nodes, enabling them to operate with lower resource overhead. Same as the transaction sharding, maintaining consistency and security across shards is complex, especially during cross-shard transactions.

These three sharding types are often combined or all of them are used simultaneously in actual sharding solutions. Since network sharding is the basis for transaction sharding and state sharding~\cite{huang2022brokerchain}, the common combinations in actual sharding schemes are:

\begin{itemize}
    \item Network and Transaction Sharding;
    
    \item Transaction and State Sharding;
    
    \item Combines all three types.
\end{itemize}

The first two combinations are often referred to as partial sharding, while combinations that use all three sharding methods are referred to as full sharding.

\section{Existing sharding solutions}

\subsection{Academic solutions}

\subsubsection{Partial sharding solutions}

\textbf{Elastico}
is the first blockchain sharding protocol~\cite{luu2016secure}, partitioning nodes into multiple committees, each responsible for processing a subset of transactions. Elastico does not support cross-shard transactions and state sharding, which means that every node maintains a full copy of the ledger~\cite{zheng2021meepo}. A directory committee coordinates the shards and aggregates the final blocks. However, this architecture introduces significant communication overhead, as each shard must frequently communicate with the directory committee and broadcast finalized blocks. In terms of security, Elastico introduces secure shard formation using Proof-of-Work (PoW) and Byzantine Fault Tolerant (BFT) consensus within each shard~\cite{liu2024dynashard}. Another partial sharding solution is \textbf{Repchain}~\cite{huang2020repchain}, which highlights differences between nodes in their activities and their role in consensus~\cite{bulgakov2024scalability}. A reputation-based system is built to assign roles and responsibilities within the network, aiming to enhance security and efficiency.

\subsubsection{Full sharding solutions}

\textbf{OmniLedger }
enhances Elastico by introducing Atomix~\cite{kokoris2018omniledger}, a two-phase commit protocol that ensures atomicity in cross-shard transactions. It uses a ByzCoin-like consensus mechanism for faster intra-shard consensus. This design reduces communication overhead compared to Elastico through more efficient management of cross-shard transactions and minimized global broadcasts. For security, OmniLedger employs randomized shard reconfiguration to thwart shard takeovers, although this added layer of complexity can potentially expand the attack surface.

\textbf{RapidChain }
~\cite{zamani2018rapidchain} introduces parallel transaction processing between shards and optimizes communication with a gossip protocol. It uses a novel identity generation mechanism to avoid trusting setup. The system significantly reduces communication overhead via lightweight intra-shard consensus and an efficient gossip-based network model. RapidChain effectively counters adaptive adversaries by utilizing robust randomness generation for shard reconfiguration, which ensures both balanced and secure shard formation.

\subsubsection{Comparison}

\textbf{Scalability. } OmniLedger and RapidChain are designed so that their transaction throughput can increase linearly with network size, exhibiting good scalability. In contrast, Elastico's scalability is limited by its ability to process transactions across shards~\cite{chen2020lightweight}.

\textbf{Communication overhead. } Elastico and OmniLedger use the BFT protocol in the intra-shard consensus process, which has a relatively high communication overhead. RepChain's communication overhead depends on the design of its reputation system, while RapidChain strives to reduce the communication overhead through optimized protocol design.

\textbf{Security. } OmniLedger and RapidChain have an advantage in security because they are designed to focus more on the randomness of the sharding selection and the secure handling of cross-shard transactions, while RepChain enhances security by introducing a reputation mechanism but needs to ensure the reliability of the reputation system itself, and Elastico's security relies mainly on the randomness of the sharding size and the allocation of nodes~\cite{hafid2020scaling}.

\subsection{Industrial solutions}

Sharding solutions within the blockchain ecosystem aim to address scalability, transaction speed, and efficiency challenges. Notable implementations include Ethereum 2.0's Danksharding, BNB Smart Chain, and Zilliqa's dynamic sharding.

\subsubsection{Ethereum}
Ethereum introduces Danksharding~\cite{danksharding}, which divides the Ethereum blockchain into smaller shards that can independently process transactions and data in parallel. A key innovation of Danksharding is the use of data availability sampling (DAS), a technique that enables lightweight nodes to verify the availability of data in a shard without the need to download the entire shard, reducing the computational burden on individual nodes. In addition, execution and consensus are separated, with execution handled by shards and consensus by the beacon chain~\cite{cassez2022formal}. Each shard has a single block proposer per epoch, which streamlines block creation. Furthermore, Danksharding incorporates Proto-Danksharding, which introduces blob-carrying transactions~\cite{park2024impact}. This innovation facilitates more efficient data storage, increasing the throughput of the network by allowing larger volumes of data to be processed in each block. Ethereum's performance is currently constrained, but is expected to enable thousands of transactions per second (TPS) with Danksharding.

\subsubsection{BNB Smart Chain}
BNB Smart Chain (BSC) addresses scalability through modular design and horizontal scaling rather than implementing traditional blockchain sharding~\cite{opBNB}. As an Ethereum Virtual Machine (EVM) compatible network, BSC enhances transaction throughput by optimizing block production within a single chain, while its multichain architecture allows for workload distribution across interconnected chains, resembling sharding in functionality. Layer-2 solutions and cross-chain protocols further facilitate distributed processing, improving efficiency. Although BSC avoids full network sharding due to EVM constraints and its centralized validator model of the Proof of Staked Authority (PoSA) consensus protocol, its focus on sidechains~\cite{wang2019sok}, cross-chain interoperability, and modularity reflects a practical approach to achieve scalability within its ecosystem.

\subsubsection{Zilliqa}
Zilliqa employs a dynamic sharding architecture that partitions its network into multiple shards~\cite{secure2018zilliqa}, each capable of independently processing transactions. These shards are formed dynamically based on network load, allowing the system to scale in real-time according to user activity or dApp demands. Security is ensured through a Practical Byzantine Fault Tolerance (PBFT) consensus mechanism~\cite{castro1999practical}, which tolerates up to a third of the faulty nodes. Zilliqa’s X-Shards architecture is operational and supports high-throughput applications in sectors such as gaming and financial services, capable of handling thousands of TPS.

\subsection{Hedera}

The Hedera is a consensus algorithm utilizing a Directed Acyclic Graph (DAG). Its Gossip about Gossip protocol enables nodes to dynamically disseminate and reference event blocks, building a DAG structure~\cite{raikwar2024sok}. The consensus is reached through double approval, where a block refers to a supermajority of blocks from the previous round. Once consensus is established, events are assigned received timestamps and sorted to ensure total ordering. This eliminates the need for mining, offering fairness, low latency, and high throughput.

Hedera utilizes the Hashgraph consensus mechanism, a unique distributed ledger technology that achieves asynchronous Byzantine Fault Tolerance (ABFT)~\cite{baird2020hashgraph}. Unlike traditional blockchain systems, Hashgraph uses a DAG structure to record transactions through a process called gossip about gossip~\cite{baird2016swirlds}. This mechanism enables nodes to exchange transaction information rapidly and efficiently, ensuring a high degree of security and fairness while minimizing computational overhead. Consensus timestamps and virtual voting further enhance the system's determinism and scalability, supporting fast finality and equitable transaction ordering.

\textbf{Gossip about Gossip. }
The Gossip about Gossip protocol is a foundational element of Hashgraph, designed to ensure efficient communication and rapid consensus in a decentralized network. This protocol enables nodes to share information dynamically by exchanging not only new transactions but also metadata about prior communications, thereby building a complete and verifiable history of events. Each node, upon receiving the data, propagates it further, creating an exponentially growing web of shared knowledge. This approach minimizes redundancy while maximizing the speed of information dissemination.

To maintain a consistent Hashgraph, the protocol incorporates mechanisms for tracking and validating event relationships through cryptographic hashes. Each event contains references to the creator's previous event and a peer's prior event, forming a secure lineage that links all events across the network. The nodes independently trace and validate these connections, ensuring that each participant reconstructs an identical and chronologically consistent Hashgraph. This process eliminates the need for mining or leader-based coordination.

\textbf{Virtual voting. }
In this process, the nodes independently determine the consensus by referencing the history of gossip events, which contain timestamps and cryptographic hashes of previous communications. By analyzing this shared history, each node can simulate the vote of others without exchanging additional messages, achieving consensus efficiently and securely.

Hedera represents cutting-edge distributed ledger technology that achieves scalability through its Gossip about Gossip protocol and virtual voting consensus mechanism. The system's reliance on events, lightweight yet robust units of transaction data, supports its scalability and resilience. However, as the Hedera  network continues to expand, the challenges of scaling storage, communication overhead, and security become increasingly significant. To address these issues, sharding emerges as a promising solution to optimize the network's architecture. Sharding enables partitioning of the ledger into smaller, parallelizable subsets, which can enhance Hedera's ability to manage increasing transaction volumes while maintaining performance and security standards.

\begin{figure*}[ht]
	\centering
	\includegraphics[scale=1]{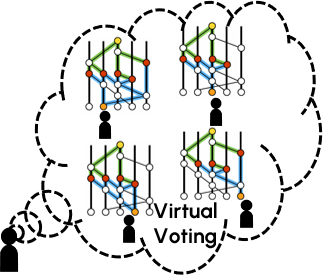}
	\caption{Virtual voting}
	\label{fig:Voting}
\end{figure*}

\section{Hedera-Specific Challenges and Solutions}
In integrating sharding into Hedera, several unique challenges must be addressed. Below, we discuss three major challenges along with corresponding potential solutions.

\subsection{Challenge 1: Integrating Sharding with the DAG-Based Consensus}
Hedera’s consensus mechanism relies on a continuously growing Hashgraph formed via the Gossip about Gossip protocol. Partitioning the network into Local and Global Committees requires segmenting this DAG, while still ensuring that all shards maintain a synchronized and verifiable global order.

\textbf{Possible Solution.} Develop a refined data partitioning model that effectively ``slices'' the DAG into interlinked subgraphs corresponding to each shard. To achieve this, each Local Committee can periodically checkpoint its local Hashgraph state to the Global Committee. By embedding cryptographic links---such as signed timestamps or hash pointers---into these checkpoints, cross-shard synchronization is maintained without sacrificing the inherent benefits of the DAG structure.

\subsection{Challenge 2: Increased Complexity in Cross-Shard Communication}
In a unified Hashgraph, nodes communicate efficiently via the Gossip Sync protocol. After implementing sharding, transactions that originate in one shard and involve nodes in another require additional communication steps through local coordinators and the Global Committee. This extra multi-stage process may lead to increased latency and higher overhead under heavy cross-shard transaction volumes.

\textbf{Possible Solution.}  To mitigate this overhead, optimize cross-shard communication by implementing batch processing and lightweight atomic commit protocols (e.g., two-phase commit) at the Global Committee level. Bundling transactions to reduce per-transaction coordination and defining standardized message formats can help minimize latency while preserving consistency across shards. Additionally, adaptive protocol adjustments based on current network load can further enhance efficiency.

\subsection{Challenge 3: Ensuring Secure Committee Reconfiguration}
 Continuous committee reorganization is required to counter security threats such as Sybil or join-leave attacks. However, ensuring that the randomized redistribution of nodes between Local and Global Committees remains secure and unpredictable is challenging. In particular, the use of consensus timestamps as randomness seeds must be carefully calibrated to avoid manipulation or predictability.

\textbf{Possible Solution.}
Employ robust randomness generation mechanisms, such as decentralized randomness beacons or verifiable delay functions (VDFs). By incorporating these techniques, node assignment during committee reconfiguration becomes truly unpredictable. Moreover, rather than relying solely on fixed thresholds for triggering reassignments, dynamic thresholds can be introduced. For example, thresholds could be adjusted in accordance with the current transaction volume or network load, so that a committee is reorganized only when a specified fraction of nodes exit relative to the prevailing network conditions. This can maintains high security and fault tolerance across both local and global layers, but also adapts to varying operational demands.

\section{New Sharding Design for Hedera}
In this section, we present overview of our proposed sharding solution for Hedera, and detail the technical designs.

\subsection{Overview}
\begin{figure*}[ht]
	\centering
	\includegraphics[width = \linewidth]{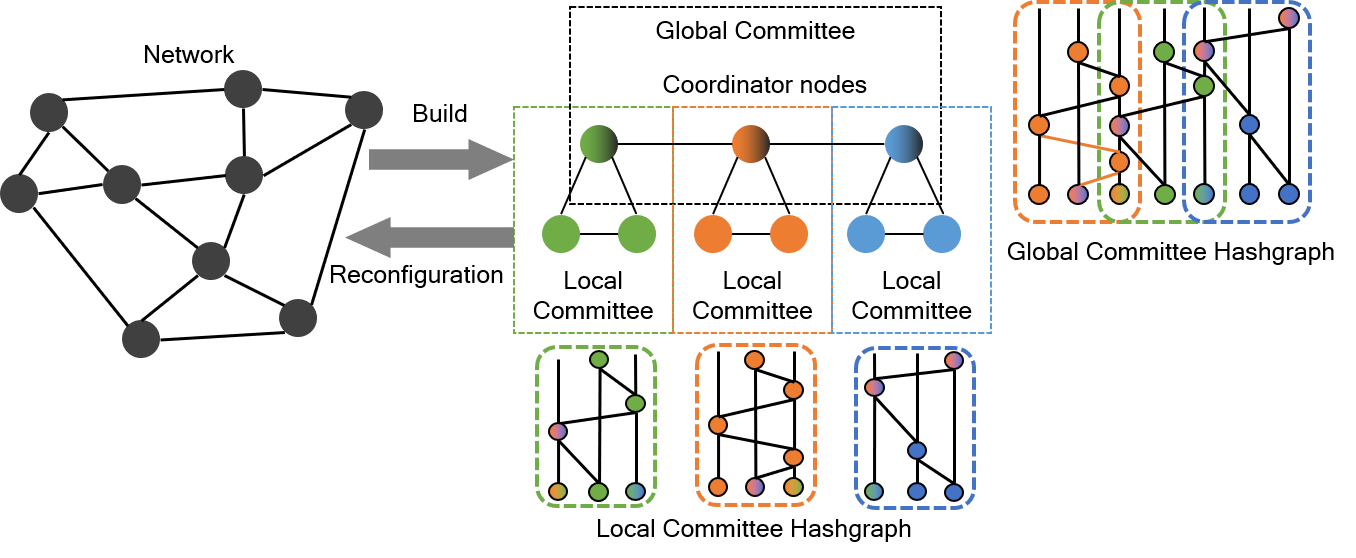}
	\caption{Design Overview}
	\label{fig:overview}
\end{figure*}

A general overview of the design of the sharding solution is shown in Fig.~\ref{fig:overview} The primary objectives of implementing sharding in Hedera include:

\begin{itemize}
\item \textbf{Reducing Storage Requirements.}
In the current architecture, each node stores the entire Hashgraph, which grows with the size of the network, leading to unsustainable storage demands. Sharding divides the network into smaller, manageable segments, enabling nodes to store only a portion of the Hashgraph, thus improving scalability.

\item \textbf{Reducing Communication Overhead.}
Hedera's Gossip Sync protocol requires extensive data sharing among nodes. Sharding limits the communication scope within localized groups, decreasing the communication cost under the same transaction throughput and enhancing network efficiency.

\item \textbf{Enhancing Security and Reliability.}
Large networks are susceptible to attacks. For the Hashgraph protocol, the cheap creation of nodes' identities and the ability to send conflicting events make the network potentially vulnerable to attacks such as Sybil attacks~\cite{douceur2002sybil} and Equivocation attacks~\cite{chun2007attested}. Sharding helps mitigate these attacks by randomly assigning nodes to shards and limits the scope and severity of the attack.

\end{itemize}

\subsection{Detailed Designs}

The sharding solution introduces a hybrid structure composed of Local Committees and a Global Committee, achieving a balance between storage optimization, efficient communication, and network security.

\subsubsection{Partitioning Nodes into Local Committees.}

Local Committees are formed by randomly assigning nodes to smaller groups, each responsible for managing a subset of the network's data and transactions. This approach reduces the storage and communication burdens on individual nodes.

Each Local Committee handles transactions relevant to its assigned shard, processed using the Hashgraph protocol to ensure consistency and correct event ordering. The intuition here is that limiting interactions to nodes within the same Local Committee significantly decreases communication overhead, streamlining the Gossip Sync process. Moreover, randomized assignment of nodes maintains security by reducing the likelihood of malicious actors controlling a shard.

\subsubsection{Establishing a Global Committee.}

Within each Local Committee, a member is randomly selected to be a coordinator. All the coordinators from different Local Committees form the Global Committee. Therefore, a coordinator is a member of both a specific Local Committee and the Global Committee. The Global Committee plays a pivotal role in maintaining network cohesion and ensuring the reliability of inter-shard operations. Its responsibilities include

\textbf{Data Redundancy.}
The Global Committee stores copies of the Hashgraphs managed by the Local Committees, ensuring that critical data is preserved even if individual shards face disruptions.

\textbf{Cross-Shard Transaction Processing.}
Handling transactions that involve multiple shards is a crucial aspect of the proposed sharding framework. The Global Committee facilitates communication between Local Committees, enabling seamless processing of cross-shard transactions and preserving data consistency across the network. The process involves:

\begin{enumerate}
    \item \textbf{Local Committee Processing:} Transactions are initially processed within their originating shard, with events containing cross-shard data being marked for further processing. For instance, Local Committee A processes events involving cross-shard transactions with Local Committee B by transmitting them to its designated coordinator via Gossip Sync. The coordinator places these cross-shard transactions in the cache queue and generates a new event that excludes the cross-shard transactions. This newly created event is then gossiped to other members within Local Committee A. Since nodes within each Local Committee lack access to the Hashgraph of other shards, they are unable to directly validate or accept events originating from different shards.
    
    \item \textbf{Global Committee Coordination:} After other members of the Global Committee complete a Gossip Sync with the coordinator of Local Committee A, the coordinator generates a new event that includes the cross-shard transactions. This event is then propagated within the Global Committee. When the coordinator of Local Committee B receives these cross-shard transactions, it adds them to its cache queue. This process ensures that cross-shard transactions are effectively synchronized across both local and global committees while maintaining consistency within the Hashgraph.

    \item \textbf{Inter-Shard Synchronization:} Finally, when members of Local Committee B engage in Gossip Sync with their coordinator, the coordinator can generate an event that incorporates the cross-shard transactions from the cache queue. This newly created event is then propagated to the other members of Local Committee B via Gossip Sync.
\end{enumerate}

Through this process, the cross-shard transactions become known and accepted by the target committee, ensuring their seamless integration into the overall system while maintaining the consistency and integrity of the Hashgraph of each shard.

\begin{figure*}[ht]
	\centering
	\includegraphics[scale=0.6]{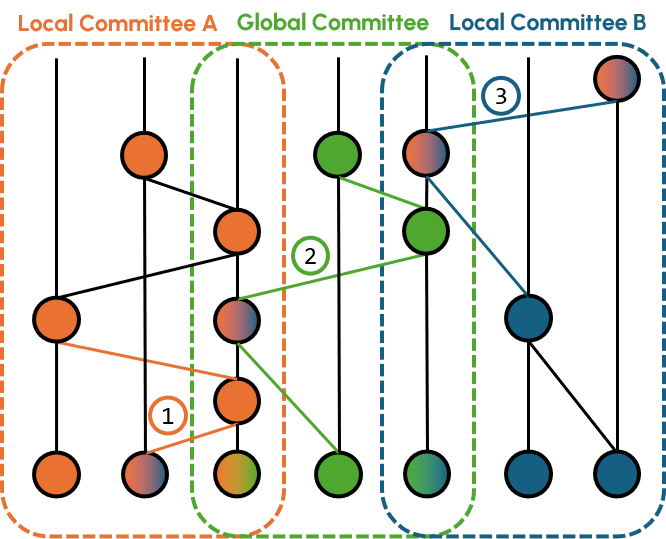}
	\caption{Cross-Shard Transaction Processing}
	\label{fig:Cross-Shard}
\end{figure*}

\textbf{Dynamic Reconfiguration of Committees.}
In committee-based systems, the leave of nodes and the addition of new nodes change the composition of committees, and an adversary may use join-leave attacks to control certain committees~\cite{awerbuch2006towards}. To improve security and ensure long-term adaptability, this solution first randomly allocates the committee to which the newly joined node belongs. The newly joined node sends a request to the Global Committee, and the node that receives the request sends a transaction, and the consensus timestamp of this transaction is used as a random number to compute the committee to which the node belongs. Second, the solution reorganizes the committee at the appropriate time. Specifically, a committee is reorganized when the number of exiting nodes in a committee exceeds half of the number of committees. The reorganization is done by the coordinator of the committee sending a committee reorganization transaction to the Global Committee, and the consensus timestamp of this transaction is used as a random number to count the other committees involved in the reorganization. The coordinator of these committees sends an intra-shard committee reorganization transaction to select the split member with its consensus timestamp. Finally, the committee's coordinator sends an intra-shard transaction to reselect the coordinator after the committee reorganization. This dynamic adjustment prevents adversaries from gaining prolonged control over any shard and maintains resilience against node failures or malicious activities.

\subsection{Implementation Challenges}

Implementing the proposed sharding solution requires managing efficient and timely cross-shard transaction communication between Local Committees and the Global Committee; given the asynchronous nature of the Hashgraph and its Gossip Sync protocol, coordinating transaction propagation and maintaining consistency across shards can introduce significant latency and increased overhead. In such an environment, ensuring that each committee correctly and quickly processes both local and cross-shard transactions is a deep technical hurdle that necessitates careful protocol tuning and network optimization.

To overcome this challenge, one promising approach is to redesign the cross-shard communication mechanism by introducing batch processing for cross-shard transactions along with lightweight atomic commit protocols (e.g., a two-phase commit). These enhancements not only reduce per-transaction coordination overhead and latency but also provide a robust framework for maintaining data consistency and security across committees.

\section{Performance Analysis}

\subsection{Efficiency}
For the communication overhead of the Hashgraph protocol, we assume an ideal situation where there is a steady and large flow of transactions in the network~\cite{baird2020hashgraph}, which makes it so that all the events generated by the nodes in the network contain at least one transaction. In a network consisting of \( n \) nodes, each node must propagate an event to the remaining \( n-1 \) nodes using a gossip protocol. Each of these \( n-1 \) gossip transmissions generates \( n-1 \) new events. Consequently, the total number of events grows exponentially with the number of communication rounds, \( k \). In the optimal scenario, each generated event contains exactly one transaction, with no empty events. Under these conditions, since the communication complexity of gossiping a single event is \( O(n) \), the network transaction throughput (\( T_{\text{throughput}} \)) is directly proportional to the bandwidth (\( B \)) of the nodes: \( T_{\text{throughput}} \propto B \).

For a network that employs the sharding solution, let the number of shards be \( s \). Under the same network transaction throughput \( T_{\text{throughput}} \), the bandwidth requirement per node in the non-sharded network is \( B \). On average, each node sends \( \frac{T_{\text{throughput}}}{n} \) events per unit of time and the size of the event that contains one transaction is \( E \). Then the communication cost per node, denoted as \( C \), is given by:  

\[ C = (n-1) \cdot \frac{T_{\text{throughput}}}{n} \cdot E. \]
After sharding, each shard handles an intra-shard transaction throughput of \( \frac{T_{\text{throughput}}}{s} \). Consequently, the bandwidth requirement per node within a shard decreases to \( \frac{B}{s} \), and the communication cost per node reduces to:  
\[ C = \left( \frac{n}{s} - 1 \right) \cdot \frac{T_{\text{throughput}}}{n} \cdot E, \]
where \( E \) represents the size of a single event.
Assuming that the global committee has sufficient bandwidth to manage the cross-shard transaction throughput \( T_{\text{cross}} \), the communication cost for a node to send cross-shard transactions is:
\(C_{\text{cross}} = T_{\text{cross}} \cdot E.\)
The cost of sending a single event remains linear in complexity.

Regarding storage size \( S \), in the ideal case, the size of the Hashgraph is proportional to the number of transactions \( T \) generated in the network. For a Local Committee, the storage consumption is reduced to \( 1/s \) of the presharding size due to the division of transactions among the shards. However, for the Global Committee, the storage consumption remains unchanged as it needs to handle all cross-shard transactions and metadata.

\subsection{Security}

It is assumed that, for any honest participants, they will eventually attempt to sync with each other and will ultimately succeed. Digital signatures and cryptographic hashes are assumed to be secure, and the system operates completely asynchronously. Under these assumptions, the Hashgraph consensus protocol is shown to satisfy the Byzantine Fault Tolerance Theorem~\cite{baird2016swirlds}. Therefore, the network can tolerate up to \( f \) faulty nodes, where \( f < 1/3 \) for \( N \) total nodes. Since the hashgraph consensus protocol is used within the Local and Global Comittee to determine the total order of events, for intra-shard transactions, the protocol achieves safety if the committee has no more than \( f < 1/3 \) fraction of corrupt nodes. For cross-shard transactions, each event created by an honest member will eventually be assigned a consensus position in the total order of events, with probability 1 if the Global Comittee and target Local Committee have no more than \( f < 1/3 \) fraction of corrupt nodes.

We would expect the adversary to disagree on a fraction \( f \) of the events, while the honest nodes agree on the remaining \( 1 - f \) fraction. However, this is not exactly what happens in the committees. The influence of the adversary increases because it is easier to increase the fraction of adversarial identities within each shard compared to the system as a whole. Therefore, the random assignment of new members to different committees and the reorganization of committees with too few members greatly diminishes the influence of adversary in these various committees, reducing the probability of a successful attack.

\subsection{Proposed Method vs. Sharding in Hedera}

Compared to the original Hedera design which primarily relies on a single main shard and a governance committee to manage the entire network and performs cross-shard transactions by randomly assigning nodes to different shards, our approach builds a dual-layer architecture that combines local committees with a global committee. In the proposed new design, each local committee is solely responsible for handling in-shard data storage and transaction processing, significantly reducing the storage and communication burden on individual nodes, while the global committee efficiently coordinates cross-shard transactions and manages data redundancy to ensure state consistency across shards; in addition, by introducing dynamic committee reorganization and batch processing for cross-shard transactions, we optimize traditional push-based cross-shard communication, enhancing the efficiency of parallel consensus as well as improving system security and resistance to attacks, which results in markedly better scalability and stability compared to the original design.

\subsection{Reliability}

Within each committee, all nodes store the complete Hashgraph of their respective committee. Additionally, nodes in the Global Committee provide redundant backups of the Hashgraphs from other committees. Consequently, the total number of copies of a complete committee Hashgraph is given by:
\(\frac{n}{s} + s - 1.\)
If an entire shard fails, it is still possible to reorganize the committee and resume work on the failed shard using an additional backup of the Hashgraph.


\section{Conclusion}


We explore the advancements, methodologies, and adoption potential of sharding in the context of Hedera, emphasizing its role as a solution to blockchain scalability challenges. By investigating various academic and industrial sharding techniques, we highlighted their unique advantages and trade-offs. Building on these insights, we proposed a sharding framework tailored to Hedera, incorporating Local and Global Committees to improve efficiency, scalability, and security. Our analysis demonstrates that the proposed solution effectively reduces storage and communication overhead. The solution's redundant storage design and retained Byzantine Fault Tolerance ensures data reliability and resistance to one-third failure nodes. The dynamic reconfiguration of the committee further enhances the security and resilience of the shards. The sharding approach thus represents a feasible pathway for integrating sharding into Hedera's architecture, fostering its adaptability to increasing network demands.

\bibliographystyle{ACM-Reference-Format}
\bibliography{references}


\begin{thebibliography}{37}


\ifx \showCODEN    \undefined \def \showCODEN     #1{\unskip}     \fi
\ifx \showDOI      \undefined \def \showDOI       #1{#1}\fi
\ifx \showISBNx    \undefined \def \showISBNx     #1{\unskip}     \fi
\ifx \showISBNxiii \undefined \def \showISBNxiii  #1{\unskip}     \fi
\ifx \showISSN     \undefined \def \showISSN      #1{\unskip}     \fi
\ifx \showLCCN     \undefined \def \showLCCN      #1{\unskip}     \fi
\ifx \shownote     \undefined \def \shownote      #1{#1}          \fi
\ifx \showarticletitle \undefined \def \showarticletitle #1{#1}   \fi
\ifx \showURL      \undefined \def \showURL       {\relax}        \fi
\providecommand\bibfield[2]{#2}
\providecommand\bibinfo[2]{#2}
\providecommand\natexlab[1]{#1}
\providecommand\showeprint[2][]{arXiv:#2}

\bibitem[Attaran(2022)]%
        {attaran2022blockchain}
\bibfield{author}{\bibinfo{person}{Mohsen Attaran}.} \bibinfo{year}{2022}\natexlab{}.
\newblock \showarticletitle{Blockchain technology in healthcare: Challenges and opportunities}.
\newblock \bibinfo{journal}{\emph{International Journal of Healthcare Management}} \bibinfo{volume}{15}, \bibinfo{number}{1} (\bibinfo{year}{2022}), \bibinfo{pages}{70--83}.
\newblock


\bibitem[Awerbuch and Scheideler(2006)]%
        {awerbuch2006towards}
\bibfield{author}{\bibinfo{person}{Baruch Awerbuch} {and} \bibinfo{person}{Christian Scheideler}.} \bibinfo{year}{2006}\natexlab{}.
\newblock \showarticletitle{Towards a scalable and robust DHT}. In \bibinfo{booktitle}{\emph{Proceedings of the eighteenth annual ACM symposium on Parallelism in algorithms and architectures}}. \bibinfo{pages}{318--327}.
\newblock


\bibitem[Bagui and Nguyen(2015)]%
        {bagui2015database}
\bibfield{author}{\bibinfo{person}{Sikha Bagui} {and} \bibinfo{person}{Loi~Tang Nguyen}.} \bibinfo{year}{2015}\natexlab{}.
\newblock \showarticletitle{Database sharding: to provide fault tolerance and scalability of big data on the cloud}.
\newblock \bibinfo{journal}{\emph{International Journal of Cloud Applications and Computing (IJCAC)}} \bibinfo{volume}{5}, \bibinfo{number}{2} (\bibinfo{year}{2015}), \bibinfo{pages}{36--52}.
\newblock


\bibitem[Baird(2016)]%
        {baird2016swirlds}
\bibfield{author}{\bibinfo{person}{Leemon Baird}.} \bibinfo{year}{2016}\natexlab{}.
\newblock \showarticletitle{The swirlds hashgraph consensus algorithm: Fair, fast, byzantine fault tolerance}.
\newblock \bibinfo{journal}{\emph{Swirlds Tech Reports SWIRLDS-TR-2016-01, Tech. Rep}}  \bibinfo{volume}{34} (\bibinfo{year}{2016}), \bibinfo{pages}{9--11}.
\newblock


\bibitem[Baird and Luykx(2020)]%
        {baird2020hashgraph}
\bibfield{author}{\bibinfo{person}{Leemon Baird} {and} \bibinfo{person}{Atul Luykx}.} \bibinfo{year}{2020}\natexlab{}.
\newblock \showarticletitle{The hashgraph protocol: Efficient asynchronous BFT for high-throughput distributed ledgers}. In \bibinfo{booktitle}{\emph{2020 International Conference on Omni-layer Intelligent Systems (COINS)}}. IEEE, \bibinfo{pages}{1--7}.
\newblock


\bibitem[BNB Chain(2024)]%
        {opBNB}
BNB Chain \bibinfo{year}{2024}\natexlab{}.
\newblock \bibinfo{title}{opBNB - High-performance layer 2 solution}.
\newblock \bibinfo{howpublished}{website}.
\newblock
\newblock
\shownote{\url{https://docs.bnbchain.org/bnb-opbnb/overview/}}.


\bibitem[Bulgakov et~al\mbox{.}(2024)]%
        {bulgakov2024scalability}
\bibfield{author}{\bibinfo{person}{Andrey~L Bulgakov}, \bibinfo{person}{Anna~V Aleshina}, \bibinfo{person}{Sergey~D Smirnov}, \bibinfo{person}{Alexey~D Demidov}, \bibinfo{person}{Maxim~A Milyutin}, {and} \bibinfo{person}{Yanliang Xin}.} \bibinfo{year}{2024}\natexlab{}.
\newblock \showarticletitle{Scalability and Security in Blockchain Networks: Evaluation of Sharding Algorithms and Prospects for Decentralized Data Storage}.
\newblock \bibinfo{journal}{\emph{Mathematics}} \bibinfo{volume}{12}, \bibinfo{number}{23} (\bibinfo{year}{2024}), \bibinfo{pages}{3860}.
\newblock


\bibitem[Cassez et~al\mbox{.}(2022)]%
        {cassez2022formal}
\bibfield{author}{\bibinfo{person}{Franck Cassez}, \bibinfo{person}{Joanne Fuller}, {and} \bibinfo{person}{Aditya Asgaonkar}.} \bibinfo{year}{2022}\natexlab{}.
\newblock \showarticletitle{Formal verification of the ethereum 2.0 beacon chain}. In \bibinfo{booktitle}{\emph{International Conference on Tools and Algorithms for the Construction and Analysis of Systems}}. Springer, \bibinfo{pages}{167--182}.
\newblock


\bibitem[Castro et~al\mbox{.}(1999)]%
        {castro1999practical}
\bibfield{author}{\bibinfo{person}{Miguel Castro}, \bibinfo{person}{Barbara Liskov}, {et~al\mbox{.}}} \bibinfo{year}{1999}\natexlab{}.
\newblock \showarticletitle{Practical byzantine fault tolerance}. In \bibinfo{booktitle}{\emph{OsDI}}, Vol.~\bibinfo{volume}{99}. \bibinfo{pages}{173--186}.
\newblock


\bibitem[Chen and Wang(2020)]%
        {chen2020lightweight}
\bibfield{author}{\bibinfo{person}{Huan Chen} {and} \bibinfo{person}{Yijie Wang}.} \bibinfo{year}{2020}\natexlab{}.
\newblock \showarticletitle{A lightweight scalable protocol for public blockchain}.
\newblock \bibinfo{journal}{\emph{J. Comput. Res. Develop.}} \bibinfo{volume}{57}, \bibinfo{number}{7} (\bibinfo{year}{2020}), \bibinfo{pages}{1555--1567}.
\newblock


\bibitem[Chun et~al\mbox{.}(2007)]%
        {chun2007attested}
\bibfield{author}{\bibinfo{person}{Byung-Gon Chun}, \bibinfo{person}{Petros Maniatis}, \bibinfo{person}{Scott Shenker}, {and} \bibinfo{person}{John Kubiatowicz}.} \bibinfo{year}{2007}\natexlab{}.
\newblock \showarticletitle{Attested append-only memory: Making adversaries stick to their word}.
\newblock \bibinfo{journal}{\emph{ACM SIGOPS Operating Systems Review}} \bibinfo{volume}{41}, \bibinfo{number}{6} (\bibinfo{year}{2007}), \bibinfo{pages}{189--204}.
\newblock


\bibitem[Dang et~al\mbox{.}(2019)]%
        {dang2019towards}
\bibfield{author}{\bibinfo{person}{Hung Dang}, \bibinfo{person}{Tien Tuan~Anh Dinh}, \bibinfo{person}{Dumitrel Loghin}, \bibinfo{person}{Ee-Chien Chang}, \bibinfo{person}{Qian Lin}, {and} \bibinfo{person}{Beng~Chin Ooi}.} \bibinfo{year}{2019}\natexlab{}.
\newblock \showarticletitle{Towards scaling blockchain systems via sharding}. In \bibinfo{booktitle}{\emph{Proceedings of the 2019 international conference on management of data}}. \bibinfo{pages}{123--140}.
\newblock


\bibitem[Dankrad(2022)]%
        {danksharding}
\bibfield{author}{\bibinfo{person}{Dankrad}.} \bibinfo{year}{2022}\natexlab{}.
\newblock \bibinfo{title}{New sharding design with tight beacon and shard block integration}.
\newblock \bibinfo{howpublished}{website}.
\newblock
\newblock
\shownote{\url{https://notes.ethereum.org/@dankrad/new_sharding}}.


\bibitem[Douceur(2002)]%
        {douceur2002sybil}
\bibfield{author}{\bibinfo{person}{John~R Douceur}.} \bibinfo{year}{2002}\natexlab{}.
\newblock \showarticletitle{The sybil attack}. In \bibinfo{booktitle}{\emph{International workshop on peer-to-peer systems}}. Springer, \bibinfo{pages}{251--260}.
\newblock


\bibitem[Dutta et~al\mbox{.}(2020)]%
        {dutta2020blockchain}
\bibfield{author}{\bibinfo{person}{Pankaj Dutta}, \bibinfo{person}{Tsan-Ming Choi}, \bibinfo{person}{Surabhi Somani}, {and} \bibinfo{person}{Richa Butala}.} \bibinfo{year}{2020}\natexlab{}.
\newblock \showarticletitle{Blockchain technology in supply chain operations: Applications, challenges and research opportunities}.
\newblock \bibinfo{journal}{\emph{Transportation research part e: Logistics and transportation review}}  \bibinfo{volume}{142} (\bibinfo{year}{2020}), \bibinfo{pages}{102067}.
\newblock


\bibitem[Gupta and Sadoghi(2021)]%
        {gupta2021blockchain}
\bibfield{author}{\bibinfo{person}{Suyash Gupta} {and} \bibinfo{person}{Mohammad Sadoghi}.} \bibinfo{year}{2021}\natexlab{}.
\newblock \showarticletitle{Blockchain transaction processing}.
\newblock \bibinfo{journal}{\emph{arXiv preprint arXiv:2107.11592}} (\bibinfo{year}{2021}).
\newblock


\bibitem[Hafid et~al\mbox{.}(2020)]%
        {hafid2020scaling}
\bibfield{author}{\bibinfo{person}{Abdelatif Hafid}, \bibinfo{person}{Abdelhakim~Senhaji Hafid}, {and} \bibinfo{person}{Mustapha Samih}.} \bibinfo{year}{2020}\natexlab{}.
\newblock \showarticletitle{Scaling blockchains: A comprehensive survey}.
\newblock \bibinfo{journal}{\emph{IEEE access}}  \bibinfo{volume}{8} (\bibinfo{year}{2020}), \bibinfo{pages}{125244--125262}.
\newblock


\bibitem[Huang et~al\mbox{.}(2020)]%
        {huang2020repchain}
\bibfield{author}{\bibinfo{person}{Chenyu Huang}, \bibinfo{person}{Zeyu Wang}, \bibinfo{person}{Huangxun Chen}, \bibinfo{person}{Qiwei Hu}, \bibinfo{person}{Qian Zhang}, \bibinfo{person}{Wei Wang}, {and} \bibinfo{person}{Xia Guan}.} \bibinfo{year}{2020}\natexlab{}.
\newblock \showarticletitle{Repchain: A reputation-based secure, fast, and high incentive blockchain system via sharding}.
\newblock \bibinfo{journal}{\emph{IEEE Internet of Things Journal}} \bibinfo{volume}{8}, \bibinfo{number}{6} (\bibinfo{year}{2020}), \bibinfo{pages}{4291--4304}.
\newblock


\bibitem[Huang et~al\mbox{.}(2022)]%
        {huang2022brokerchain}
\bibfield{author}{\bibinfo{person}{Huawei Huang}, \bibinfo{person}{Xiaowen Peng}, \bibinfo{person}{Jianzhou Zhan}, \bibinfo{person}{Shenyang Zhang}, \bibinfo{person}{Yue Lin}, \bibinfo{person}{Zibin Zheng}, {and} \bibinfo{person}{Song Guo}.} \bibinfo{year}{2022}\natexlab{}.
\newblock \showarticletitle{Brokerchain: A cross-shard blockchain protocol for account/balance-based state sharding}. In \bibinfo{booktitle}{\emph{IEEE INFOCOM 2022-IEEE Conference on Computer Communications}}. IEEE, \bibinfo{pages}{1968--1977}.
\newblock


\bibitem[Kokoris-Kogias et~al\mbox{.}(2018)]%
        {kokoris2018omniledger}
\bibfield{author}{\bibinfo{person}{Eleftherios Kokoris-Kogias}, \bibinfo{person}{Philipp Jovanovic}, \bibinfo{person}{Linus Gasser}, \bibinfo{person}{Nicolas Gailly}, \bibinfo{person}{Ewa Syta}, {and} \bibinfo{person}{Bryan Ford}.} \bibinfo{year}{2018}\natexlab{}.
\newblock \showarticletitle{Omniledger: A secure, scale-out, decentralized ledger via sharding}. In \bibinfo{booktitle}{\emph{2018 IEEE symposium on security and privacy (SP)}}. IEEE, \bibinfo{pages}{583--598}.
\newblock


\bibitem[Li et~al\mbox{.}(2022)]%
        {li2022achieving}
\bibfield{author}{\bibinfo{person}{Canlin Li}, \bibinfo{person}{Huawei Huang}, \bibinfo{person}{Yetong Zhao}, \bibinfo{person}{Xiaowen Peng}, \bibinfo{person}{Ruijie Yang}, \bibinfo{person}{Zibin Zheng}, {and} \bibinfo{person}{Song Guo}.} \bibinfo{year}{2022}\natexlab{}.
\newblock \showarticletitle{Achieving scalability and load balance across blockchain shards for state sharding}. In \bibinfo{booktitle}{\emph{2022 41st International Symposium on Reliable Distributed Systems (SRDS)}}. IEEE, \bibinfo{pages}{284--294}.
\newblock


\bibitem[Li et~al\mbox{.}(2023)]%
        {li2023survey}
\bibfield{author}{\bibinfo{person}{Yi Li}, \bibinfo{person}{Jinsong Wang}, {and} \bibinfo{person}{Hongwei Zhang}.} \bibinfo{year}{2023}\natexlab{}.
\newblock \showarticletitle{A survey of state-of-the-art sharding blockchains: Models, components, and attack surfaces}.
\newblock \bibinfo{journal}{\emph{Journal of Network and Computer Applications}}  \bibinfo{volume}{217} (\bibinfo{year}{2023}), \bibinfo{pages}{103686}.
\newblock


\bibitem[Liu et~al\mbox{.}(2024)]%
        {liu2024dynashard}
\bibfield{author}{\bibinfo{person}{Ao Liu}, \bibinfo{person}{Jing Chen}, \bibinfo{person}{Kun He}, \bibinfo{person}{Ruiying Du}, \bibinfo{person}{Jiahua Xu}, \bibinfo{person}{Cong Wu}, \bibinfo{person}{Yebo Feng}, \bibinfo{person}{Teng Li}, {and} \bibinfo{person}{Jianfeng Ma}.} \bibinfo{year}{2024}\natexlab{}.
\newblock \showarticletitle{DYNASHARD: Secure and Adaptive Blockchain Sharding Protocol With Hybrid Consensus and Dynamic Shard Management}.
\newblock \bibinfo{journal}{\emph{IEEE Internet of Things Journal}} (\bibinfo{year}{2024}).
\newblock


\bibitem[Liu et~al\mbox{.}(2020)]%
        {liu2020secure}
\bibfield{author}{\bibinfo{person}{Yizhong Liu}, \bibinfo{person}{Jianwei Liu}, \bibinfo{person}{Yiming Hei}, \bibinfo{person}{Wei Tan}, {and} \bibinfo{person}{Qianhong Wu}.} \bibinfo{year}{2020}\natexlab{}.
\newblock \showarticletitle{A secure shard reconfiguration protocol for sharding blockchains without a randomness}. In \bibinfo{booktitle}{\emph{2020 IEEE 19th International Conference on Trust, Security and Privacy in Computing and Communications (TrustCom)}}. IEEE, \bibinfo{pages}{1012--1019}.
\newblock


\bibitem[Liu et~al\mbox{.}(2022)]%
        {liu2022building}
\bibfield{author}{\bibinfo{person}{Yizhong Liu}, \bibinfo{person}{Jianwei Liu}, \bibinfo{person}{Marcos Antonio~Vaz Salles}, \bibinfo{person}{Zongyang Zhang}, \bibinfo{person}{Tong Li}, \bibinfo{person}{Bin Hu}, \bibinfo{person}{Fritz Henglein}, {and} \bibinfo{person}{Rongxing Lu}.} \bibinfo{year}{2022}\natexlab{}.
\newblock \showarticletitle{Building blocks of sharding blockchain systems: Concepts, approaches, and open problems}.
\newblock \bibinfo{journal}{\emph{Computer Science Review}}  \bibinfo{volume}{46} (\bibinfo{year}{2022}), \bibinfo{pages}{100513}.
\newblock


\bibitem[Luu et~al\mbox{.}(2016)]%
        {luu2016secure}
\bibfield{author}{\bibinfo{person}{Loi Luu}, \bibinfo{person}{Viswesh Narayanan}, \bibinfo{person}{Chaodong Zheng}, \bibinfo{person}{Kunal Baweja}, \bibinfo{person}{Seth Gilbert}, {and} \bibinfo{person}{Prateek Saxena}.} \bibinfo{year}{2016}\natexlab{}.
\newblock \showarticletitle{A secure sharding protocol for open blockchains}. In \bibinfo{booktitle}{\emph{Proceedings of the 2016 ACM SIGSAC conference on computer and communications security}}. \bibinfo{pages}{17--30}.
\newblock


\bibitem[Park et~al\mbox{.}(2024)]%
        {park2024impact}
\bibfield{author}{\bibinfo{person}{Seongwan Park}, \bibinfo{person}{Bosul Mun}, \bibinfo{person}{Seungyun Lee}, \bibinfo{person}{Woojin Jeong}, \bibinfo{person}{Jaewook Lee}, \bibinfo{person}{Hyeonsang Eom}, {and} \bibinfo{person}{Huisu Jang}.} \bibinfo{year}{2024}\natexlab{}.
\newblock \showarticletitle{Impact of EIP-4844 on Ethereum: Consensus Security, Ethereum Usage, Rollup Transaction Dynamics, and Blob Gas Fee Markets}.
\newblock \bibinfo{journal}{\emph{arXiv preprint arXiv:2405.03183}} (\bibinfo{year}{2024}).
\newblock


\bibitem[Raikwar et~al\mbox{.}(2024)]%
        {raikwar2024sok}
\bibfield{author}{\bibinfo{person}{Mayank Raikwar}, \bibinfo{person}{Nikita Polyanskii}, {and} \bibinfo{person}{Sebastian M{\"u}ller}.} \bibinfo{year}{2024}\natexlab{}.
\newblock \showarticletitle{SoK: DAG-based Consensus Protocols}. In \bibinfo{booktitle}{\emph{2024 IEEE International Conference on Blockchain and Cryptocurrency (ICBC)}}. IEEE, \bibinfo{pages}{1--18}.
\newblock


\bibitem[Secure(2018)]%
        {secure2018zilliqa}
\bibfield{author}{\bibinfo{person}{A Secure}.} \bibinfo{year}{2018}\natexlab{}.
\newblock \showarticletitle{The zilliqa project: A secure, scalable blockchain platform}.
\newblock  (\bibinfo{year}{2018}).
\newblock


\bibitem[Tao et~al\mbox{.}(2024)]%
        {tao2024throughput}
\bibfield{author}{\bibinfo{person}{Liping Tao}, \bibinfo{person}{Yang Lu}, \bibinfo{person}{Yuqi Fan}, \bibinfo{person}{Lei Shi}, \bibinfo{person}{Chee~Wei Tan}, {and} \bibinfo{person}{Zhen Wei}.} \bibinfo{year}{2024}\natexlab{}.
\newblock \showarticletitle{Throughput-Scalable Shard Reorganization Tailored to Node Relations in Sharding Blockchain Networks}.
\newblock \bibinfo{journal}{\emph{IEEE Transactions on Computational Social Systems}} (\bibinfo{year}{2024}).
\newblock


\bibitem[Varma(2019)]%
        {varma2019blockchain}
\bibfield{author}{\bibinfo{person}{Jayanth~Rama Varma}.} \bibinfo{year}{2019}\natexlab{}.
\newblock \showarticletitle{Blockchain in finance}.
\newblock \bibinfo{journal}{\emph{Vikalpa}} \bibinfo{volume}{44}, \bibinfo{number}{1} (\bibinfo{year}{2019}), \bibinfo{pages}{1--11}.
\newblock


\bibitem[Wang et~al\mbox{.}(2019)]%
        {wang2019sok}
\bibfield{author}{\bibinfo{person}{Gang Wang}, \bibinfo{person}{Zhijie~Jerry Shi}, \bibinfo{person}{Mark Nixon}, {and} \bibinfo{person}{Song Han}.} \bibinfo{year}{2019}\natexlab{}.
\newblock \showarticletitle{Sok: Sharding on blockchain}. In \bibinfo{booktitle}{\emph{Proceedings of the 1st ACM Conference on Advances in Financial Technologies}}. \bibinfo{pages}{41--61}.
\newblock


\bibitem[Werth et~al\mbox{.}(2023)]%
        {werth2023review}
\bibfield{author}{\bibinfo{person}{Jan Werth}, \bibinfo{person}{Mohammad~Hajian Berenjestanaki}, \bibinfo{person}{Hamid~R Barzegar}, \bibinfo{person}{Nabil El~Ioini}, {and} \bibinfo{person}{Claus Pahl}.} \bibinfo{year}{2023}\natexlab{}.
\newblock \showarticletitle{A Review of Blockchain Platforms Based on the Scalability, Security and Decentralization Trilemma.}
\newblock \bibinfo{journal}{\emph{ICEIS (1)}} (\bibinfo{year}{2023}), \bibinfo{pages}{146--155}.
\newblock


\bibitem[Xu et~al\mbox{.}(2020)]%
        {xu2020flexible}
\bibfield{author}{\bibinfo{person}{Yibin Xu}, \bibinfo{person}{Yangyu Huang}, \bibinfo{person}{Jianhua Shao}, {and} \bibinfo{person}{George Theodorakopoulos}.} \bibinfo{year}{2020}\natexlab{}.
\newblock \showarticletitle{A flexible n/2 adversary node resistant and halting recoverable blockchain sharding protocol}.
\newblock \bibinfo{journal}{\emph{Concurrency and Computation: Practice and Experience}} \bibinfo{volume}{32}, \bibinfo{number}{19} (\bibinfo{year}{2020}), \bibinfo{pages}{e5773}.
\newblock


\bibitem[Yang et~al\mbox{.}(2020)]%
        {yang2020review}
\bibfield{author}{\bibinfo{person}{Di Yang}, \bibinfo{person}{Chengnian Long}, \bibinfo{person}{Han Xu}, {and} \bibinfo{person}{Shaoliang Peng}.} \bibinfo{year}{2020}\natexlab{}.
\newblock \showarticletitle{A review on scalability of blockchain}. In \bibinfo{booktitle}{\emph{Proceedings of the 2020 2nd International Conference on Blockchain Technology}}. \bibinfo{pages}{1--6}.
\newblock


\bibitem[Zamani et~al\mbox{.}(2018)]%
        {zamani2018rapidchain}
\bibfield{author}{\bibinfo{person}{Mahdi Zamani}, \bibinfo{person}{Mahnush Movahedi}, {and} \bibinfo{person}{Mariana Raykova}.} \bibinfo{year}{2018}\natexlab{}.
\newblock \showarticletitle{Rapidchain: Scaling blockchain via full sharding}. In \bibinfo{booktitle}{\emph{Proceedings of the 2018 ACM SIGSAC conference on computer and communications security}}. \bibinfo{pages}{931--948}.
\newblock


\bibitem[Zheng et~al\mbox{.}(2021)]%
        {zheng2021meepo}
\bibfield{author}{\bibinfo{person}{Peilin Zheng}, \bibinfo{person}{Quanqing Xu}, \bibinfo{person}{Zibin Zheng}, \bibinfo{person}{Zhiyuan Zhou}, \bibinfo{person}{Ying Yan}, {and} \bibinfo{person}{Hui Zhang}.} \bibinfo{year}{2021}\natexlab{}.
\newblock \showarticletitle{Meepo: Sharded consortium blockchain}. In \bibinfo{booktitle}{\emph{2021 IEEE 37th International Conference on Data Engineering (ICDE)}}. IEEE, \bibinfo{pages}{1847--1852}.
\newblock


\end{thebibliography}

\end{document}